\documentclass{article}

\usepackage{microtype}
\usepackage{graphicx}
\usepackage{subfigure}
\usepackage{booktabs} 
\usepackage{multicol}
\usepackage{amsmath}
\usepackage{amsfonts}
\usepackage{bm}
\usepackage{bbm}
\usepackage{hyperref}
\usepackage[accepted]{icml2021}

\icmltitlerunning{Inferring Black Hole Properties from Astronomical Multivariate Time Series with Bayesian Attentive Neural Processes}

\begin{document}

\twocolumn[
\icmltitle{Inferring Black Hole Properties from Astronomical Multivariate Time Series with Bayesian Attentive Neural Processes}




\begin{icmlauthorlist}
\icmlauthor{Ji Won Park}{stanford,slac}
\icmlauthor{Ashley Villar}{columbia,cca}
\icmlauthor{Yin Li}{cca}
\icmlauthor{Yan-Fei Jiang}{cca}
\icmlauthor{Shirley Ho}{cca,princeton,nyu,cmu}
\icmlauthor{Joshua Yao-Yu Lin}{uiuc}
\icmlauthor{Phil Marshall}{stanford,slac}
\icmlauthor{Aaron Roodman}{stanford,slac}
\end{icmlauthorlist}

\icmlaffiliation{stanford}{Kavli Institute for Particle Astrophysics and Cosmology, Department of Physics, Stanford University, Stanford, CA, USA}
\icmlaffiliation{slac}{SLAC National Accelerator Laboratory, Menlo Park, CA, USA}
\icmlaffiliation{cca}{Flatiron Institute, New York, NY, USA}
\icmlaffiliation{columbia}{Department of Astronomy, Columbia University, New York, NY, USA}
\icmlaffiliation{uiuc}{University of Illinois at Urbana-Champaign, Champaign, IL, USA}
\icmlaffiliation{princeton}{Princeton University, Princeton,NJ 08540}
\icmlaffiliation{nyu}{New York university, New York, NY 10010}
\icmlaffiliation{cmu}{Carnegie Mellon University, Pittsburgh, PA 15289}

\icmlcorrespondingauthor{Ji Won Park}{jwp@stanford.edu}

\icmlkeywords{Machine Learning, ICML}

\vskip 0.3in
]



\printAffiliationsAndNotice{}  

\begin{abstract}
Among the most extreme objects in the Universe, active galactic nuclei (AGN) are luminous centers of galaxies where a black hole feeds on surrounding matter. The variability patterns of the light emitted by an AGN contain information about the physical properties of the underlying black hole. Upcoming telescopes will observe over 100 million AGN in multiple broadband wavelengths, yielding a large sample of multivariate time series with long gaps and irregular sampling. We present a method that reconstructs the AGN time series and simultaneously infers the posterior probability density distribution (PDF) over the physical quantities of the black hole, including its mass and luminosity. We apply this method to a simulated dataset of 11,000 AGN and report precision and accuracy of 0.4 dex and 0.3 dex in the inferred black hole mass. This work is the first to address probabilistic time series reconstruction and parameter inference for AGN in an end-to-end fashion.  

\end{abstract}

\section{Introduction}
\label{sec:intro}
Supermassive black holes (BHs) reside at the centers of most galaxies, feeding on diffuse matter around them. Radiation from matter falling into their gravitational pull makes these galactic centers, called active galactic nuclei (AGN), some of the most luminous in the Universe. The time variability patterns of AGN light are correlated with the physical properties of the underlying BH, such as the mass, rate of matter inflow, and age \cite{wold2008black, simm2016pan, macleod2010modeling, suberlak2021improving}.

Being so luminous, AGN can be observed out to great distances, close to the edge of the observable Universe (e.g., \citealt{mortlock2011luminous}). Characterizing faraway BHs gives us a glimpse into the little-known early Universe. By inferring BH physics from AGN light, we can gain an understanding the origin and evolution of the cosmos, including the nature of dark energy and dark matter \citep{khadka2020using}.

Upcoming large-sky telescope surveys herald an unprecedented increase in the AGN data volume. The Vera Rubin Observatory Legacy Survey of Space and Time (LSST) is projected to yield 100 million AGN time series in six optical broadband filters over ten years \cite{abell2009lsst}. Traditional methods of estimating BH properties, however, rely on expensive spectroscopic data, i.e. measurements of the AGN light at continuous wavelengths. Obtaining spectroscopy for millions of objects would be unfeasible. To take advantage of all the new data, we require an efficient method that can estimate desired quantities directly from the 6-filter LSST time series. The learned relationship could improve our understanding of BH physics, particularly as a physical model of AGN variability does not exist.

We present a method that simultaneously reconstructs the multivariate AGN time series from limited observations and infers the full posterior probability density distribution (PDF) over key BH properties. Our method is designed for irregular, multivariate time series, as telescope data often suffer from long seasonal gaps and irregular sampling as well as noise from the Earth's atmosphere and telescope optics. Uncertainty quantification is essential to optimize follow up strategies.

This work is the first deep learning pipeline that simultaneously addresses AGN light curve reconstruction and parameter inference in a probabilistic manner. It is additionally the first designed for multivariate time series. At the core of our pipeline is an attentive neural process \cite{kim2019attentive}, a type of latent variable model, that has been modified for density estimation. In the past, autoencoders have been applied to output point estimates of the unobserved portions of the light curve in a single-filter setting \cite{tachibana2020deep}. Convolutional neural nets were trained to point-estimate the redshifts based on multi-filter light curves from the Sloan Digital Sky Survey (SDSS) \cite{schneider2010sloan, pasquet2018deep}. Summary statistics from the SDSS light curves were also fed into a neural net for point-estimating BH mass and redshift \cite{lin2020agnet}. 
\vspace{-0.125in}
\section{Data}

\subsection{Multi-filter time series}
\label{sec:time-series}
The training set consisted of 11,000 simulated multi-filter time series and the corresponding target labels. The validation set contained 50 held-out examples, drawn from the same distribution as the training set.
The input was a simulated multivariate time series of the AGN flux, or brightness, in the six bandpass filters $ugrizY$. This six-dimensional light curve followed the Ornstein-Uhlenbeck (OU) process, a widely adopted stochastic model of AGN flux variability \cite{kelly2009variations}. The light curves were sampled at irregular times, with long gaps, to simulate LSST-like observations \cite{reuter2016opsim}. We added astrophysical noise to the light curves on the fly during training \citep{kessler2019models}.
\vspace{-0.125in}
\subsection{Target quantities}
\label{sec:target-quantities}
The target quantities were taken from the Second Data Challenge (DC2), a simulated LSST-like catalog \citep{LSSTDESC_dc2_paper, LSSTDESC_dc2_release_note}. For each of the $ugrizY$ filters, there were three variability parameters: the average flux ($m$), long-term amplitude ($SF_{\infty}$) of fluctuations, and characteristic timescale ($\tau$) of fluctuations. The $SF_\infty$ and $\tau$ parameters can approximately be interpreted as the maximum amplitude of flux fluctuations and the timescale to reach such an amplitude, respectively. We also included the BH mass, redshift (a measure of distance), and the i-band absolute magnitude ($M_i$; a measure of the intrinsic luminosity of the AGN). The 21 quantities in total shared correlations empirically modeled after a well-known dataset (SDSS) of quasars \cite{kelly2009variations, macleod2010modeling}.

\begin{figure}[ht]
\begin{center}
\centerline{\includegraphics[width=\columnwidth]{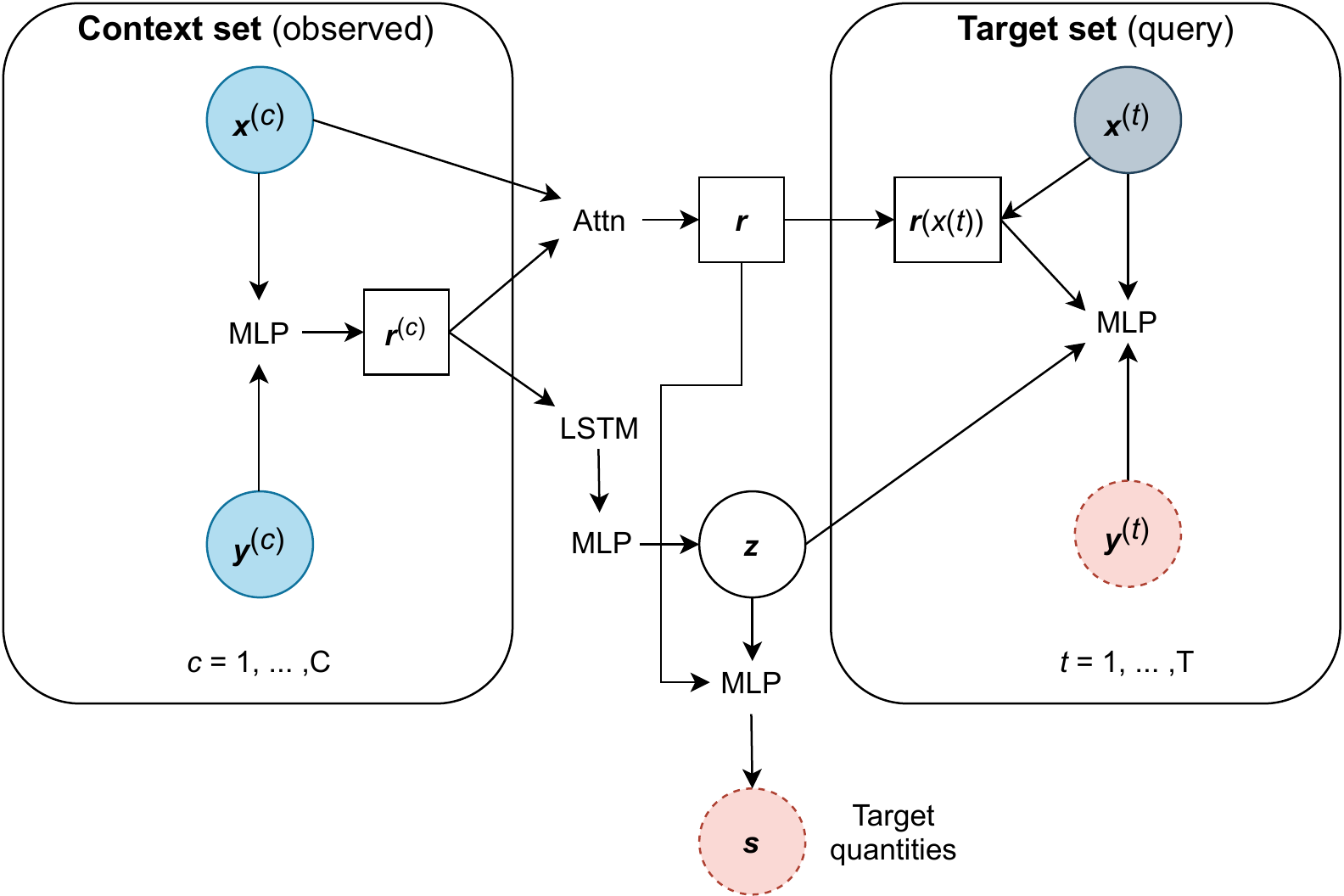}}
\caption{Model architecture of the Bayesian ANP, modified from the visualization in \cite{dubois2020npf}.}
\label{fig:diagram}
\end{center}
\vskip -0.4in
\end{figure}

\section{Method}
\subsection{Model}
See Figure \ref{fig:diagram} for a diagram of our model architecture. We used a latent attentive neural process (ANP) \cite{kim2019attentive} to model the conditional distribution over regression functions mapping our input times, $\bm{x}^{(c)} \in \mathbb{R}$, to our output $ugrizY$ fluxes, $\bm{y}^{(c)} \in \mathbb{R}^{6}$. Conditioning on observed time/flux pairs, $\left( \bm{x}_C, \bm{y}_C \right) := \left(\bm{x}^{(c)}, \bm{y}^{(c)}\right)_{c \in C}$, called the \textit{context} set, we query the model for the fluxes at some unobserved times, $\left( \bm{x}_T, \bm{y}_T \right) := \left(\bm{x}^{(t)}, \bm{y}^{(c)}\right)_{c \in C}$, called the \textit{target} set. A common convention, used here, is to set $C \subset T$.

In the \textit{deterministic} path of the forward model, each time/flux pair $\bm{x}^{(c)}, \bm{y}^{(c)}$ passes through an MLP to form a representation $\bm{r}^{(c)} \in \mathbb{R}^{h}$. Self-attention is applied to the resulting context representations and the model attends to these via cross-attention with $\bm{x}^{(t)}$ to predict $\bm{y}^{(t)}$. The attention mechanism allows it to attribute higher relative importance to time samples that are more informative for prediction. This is useful for astronomical data, where observations tend to be clustered together and some carry higher signal-to-noise than others.

The latent path encodes a global understanding of the entire time series. The encodings $\bm{r}_C := \left( \bm{r}^{(c)}\right)_{c \in C}$ are aggregated using an LSTM and passed through a multi-layer perceptron (MLP) to form a latent variable, $\bm{z}$.\footnote{The ANP, as originally proposed by Kim et al. \citeyear{kim2019attentive}, uses mean aggregation but we find that the LSTM yield better predictions of our target quantities.} We model $\bm{z}$ as a factorized Gaussian, with a Gaussian prior. Each sample of $\bm{z}$ represents one realization of the data-generating stochastic process, so $\bm{z}$ stores information about uncertainty in the predictions $\bm{y}_T$. 

As an update to the original ANP architecture, we additionally use $\bm{z}$ and the mean of the attention-reweighted vector $\bm{r}$ over the times to infer the global target quantities $\bm{s} \in \mathbb{R}^{21}$, described in Section \ref{sec:target-quantities}. An MLP takes in $\bm{z}$ and outputs the parameters defining the posterior PDF over $\bm{s}$.
\vspace{-0.1in}
\subsection{Uncertainty quantification}
We adapted the standard ANP into a Bayesian ANP (BANP) to enable posterior inference over the network weights \citep{denker1991transforming}. Our posterior on $\bm{y}_T, \bm{s}$ can be written as:
\vspace{-0.1in}
\begin{align} \label{eq:predictive_distribution}
    & p(\bm{y}_T, \bm{s} | \bm{x}_T, \bm{x}_C, \bm{y}_C) \nonumber  \\
    &= \int p(\bm{y}_T, \bm{s} | \bm{x}_T, \bm{x}_C, \bm{y}_C, W) p(W|\bm{x}_C, \bm{y}_C) \ dW,
\end{align}
where $W$ denotes the network weights. The likelihood $p(\bm{y}_T, \bm{s} | \bm{x}_T, \bm{x}_C, \bm{y}_C, W)$ captures the \textit{aleatoric} uncertainty, which exists due to the intrinsic randomness in the data-generating process. We marginalize over the weight posterior $p(W|\bm{x}_C, \bm{y}_C)$ to account for the \textit{epistemic} uncertainty, which originates from incomplete knowledge, e.g. limited training data.

\textbf{Aleatoric uncertainty: } For predicting $\bm{y}_T$, we would maximize the latent neural process likelihood \cite{garnelo2018neural}:
\begin{align} \label{eq:anp_elbo}
    & p\left(\bm{y}_T | \bm{x}_T, \bm{x}_C, \bm{y}_C, W \right) \nonumber \\
    &= \int p\left(\bm{y}_T | \bm{x}_T, \bm{r}_C, \bm{z}, W \right) q\left(\bm{z} | \bm{x}_C, \bm{y}_C, W \right) d\bm{z},
\end{align}
where we have explicitly conditioned on $W$. Similarly, the likelihood of $\bm{s}$ is
\begin{align} \label{eq:param_regressor_elbo}
    & p\left(\bm{s} | \bm{x}_C, \bm{y}_C, W \right) =\int p\left(\bm{s} | \bm{z}, W \right) q\left(\bm{z} | \bm{x}_C, \bm{y}_C, W \right) d\bm{z}.
\end{align}
We chose $p(\bm{s} | \bm{z})$ to be multivariate Gaussian with a full covariance matrix, so as to model physical correlations between the target quantities.

The total aleatoric portion of our likelihood combines Equations \ref{eq:anp_elbo} and \ref{eq:param_regressor_elbo}:
\begin{align} \label{eq:total_likelihood}
    & \log p(\bm{y}_T, \bm{s} | \bm{x}_T, \bm{x}_C, \bm{y}_C, W) \nonumber \\
    &\propto \alpha \log p\left(\bm{y}_T | \bm{x}_T, \bm{x}_C, \bm{y}_C, W \right) +  \log p\left(\bm{s} | \bm{x}_C, \bm{y}_C, W \right),    
\end{align}
where $\alpha$ is a hyperparameter controlling the relative weight between light curve reconstruction and parameter inference. 

\textbf{Epistemic uncertainty: } We used Monte Carlo (MC) dropout \citep{gal2016dropout, kendall2017uncertainties} to compute the integral in Equation \ref{eq:predictive_distribution}. MC dropout replaces the true weight posterior with a simple Bernoulli variational approximation $q(W | \bm{x}_C, \bm{y}_C)$, which can be implemented by setting random network weights to zero during training and testing. We treated the dropout rate as a hyperparameter.

\subsection{Optimization}
\textbf{Context-target split: } We provided as context the observed portions of the light curve according to the 10-year LSST observing strategy, which consisted of 500-1,000 samplings per filter. All six filters were observed at the same times. The target was the union of the context set and the light curve at every 10 days, i.e. 365 samplings per filter.

\textbf{Loss function: } The exact form of Equation \ref{eq:total_likelihood} is intractable. The evidence lower bound (ELBO) can be optimized, however, using the reparameterization trick \citep{kingma2013auto, rezende2014stochastic}:
\begin{align}
    & \log p(\bm{y}_T, \bm{s} | \bm{x}_T, \bm{x}_C, \bm{y}_C, W) \geq \nonumber \\
    & \mathbb{E}_{q(\bm{z}|\bm{x}_T, \bm{y}_T)} \left[ \alpha' \log  p(\bm{y}_T | \bm{x}_T, \bm{r}_C, \bm{z}, W) + \log p(\bm{s} | \bm{z}, W) \right] \nonumber \\
    & - D_{\rm KL} \left( q(\bm{z} | \bm{x}_T, \bm{y}_T, W) || q(\bm{z} | \bm{x}_C, \bm{y}_C, W) \right) := \mathcal{F}_{\rm ELBO}(W)
\end{align}
where $\alpha'$ serves the same reweighting purpose as $\alpha$ in Equation \ref{eq:total_likelihood}. We had $\alpha' = 10$. In each training iteration, given a realization of weights from MC dropout, $\hat W \sim q(\hat W | \bm{x}_C, \bm{y}_C)$, we performed gradient descent on $\mathcal{F}_{\rm ELBO}(\hat W)$ with a weight decay of 1e-5 using the \texttt{ADAM} optimizer \citep{kingma2014adam}. Training was done for 300 epochs in batch sizes of 40. The learning rate began with 1e-3 and was halved whenever the validation loss did not decrease for 20 epochs.

\section{Results} \label{sec:results}
\begin{figure}[ht]
\begin{center}
\centerline{\includegraphics[width=\columnwidth]{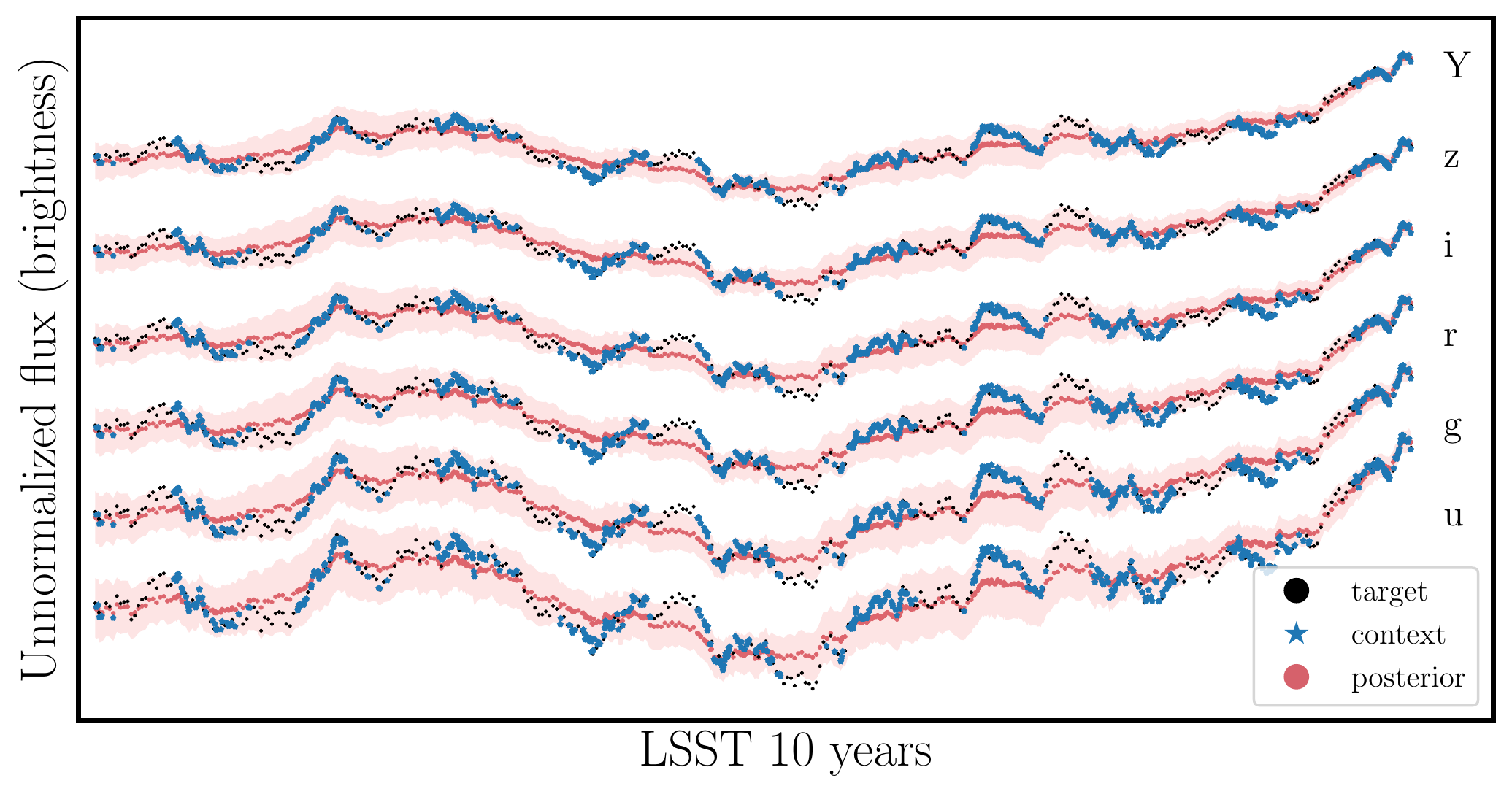}}
\caption{Reconstructed $ugrizY$ light curve for a validation AGN. Blue stars are observed \textit{context} points, provided to the network at test time. Black dots are query \textit{target} points, for which the network produces the red posteriors. The true observations lie within the 1-$\sigma$ region of our predicted light curves.}
\label{fig:light-curve}
\end{center}
\vskip -0.3in
\end{figure}
\subsection{Light curve reconstruction} \label{sec:results-reconstruction}
Figure \ref{fig:light-curve} shows the BANP posterior for a validation $ugrizY$ light curve. Our model closely reconstructs the long-term amplitude and timescale of fluctuations. This performance indicates a good understanding of $SF_\infty$ and $\tau$. While the truth is consistent with its 1-$\sigma$ credible interval at all times, the uncertainties are overestimated and the model does a poorer job of predicting the shorter-term fluctuations. One possibility is that the reconstruction loss competes with the parameter inference loss, as the model does not have to get the shorter-term fluctuations correct to predict $SF_\infty$ and $\tau$.  
\begin{table}[t]
\caption{Validation metrics of time series reconstruction and parameter recovery. Lower is better. Error bars indicate the standard deviations over three training random seeds.}
\label{metrics}
\vskip -0.3in
\begin{center}
\begin{small}
\begin{sc}
\begin{tabular}{lcccc}
\toprule
Model & \multicolumn{2}{c}{Reconstruction} & \multicolumn{2}{c}{Parameter} \\
{} &  {1-$\sigma$} & {MAE} & {1-$\sigma$} & {MAE} \\
\midrule
BANP    & {$3.1 \pm 0.9$} & {$2.2 \pm 0.8$} & {$\mathbf{51 \pm 2}$} & {$\mathbf{46 \pm 1}$}  \\
Baseline & {N/A} &  {N/A} & {$89 \pm 3$} & {$97 \pm 3$} \\
\bottomrule
\end{tabular}
\end{sc}
\end{small}
\end{center}
\vspace{-0.3in}
\end{table}

To assess the precision of reconstruction, we draw 100 light curves from our flux posterior and approximate the 1-$\sigma$ credible width at each target time as the standard deviation across the samples. We then take the average across all target times and 50 validation AGN. Similarly, to assess the accuracy, we obtain the absolute error of the central flux prediction at each target time and take the average across all target times in the validation set (mean absolute error; MAE). The resulting values, which carry units of magnitude, are listed in Table \ref{metrics}.

\subsection{Retrieval of target quantities}
\begin{figure}[ht]
\vskip -0.1in
\begin{center}
\centerline{\includegraphics[width=\columnwidth]{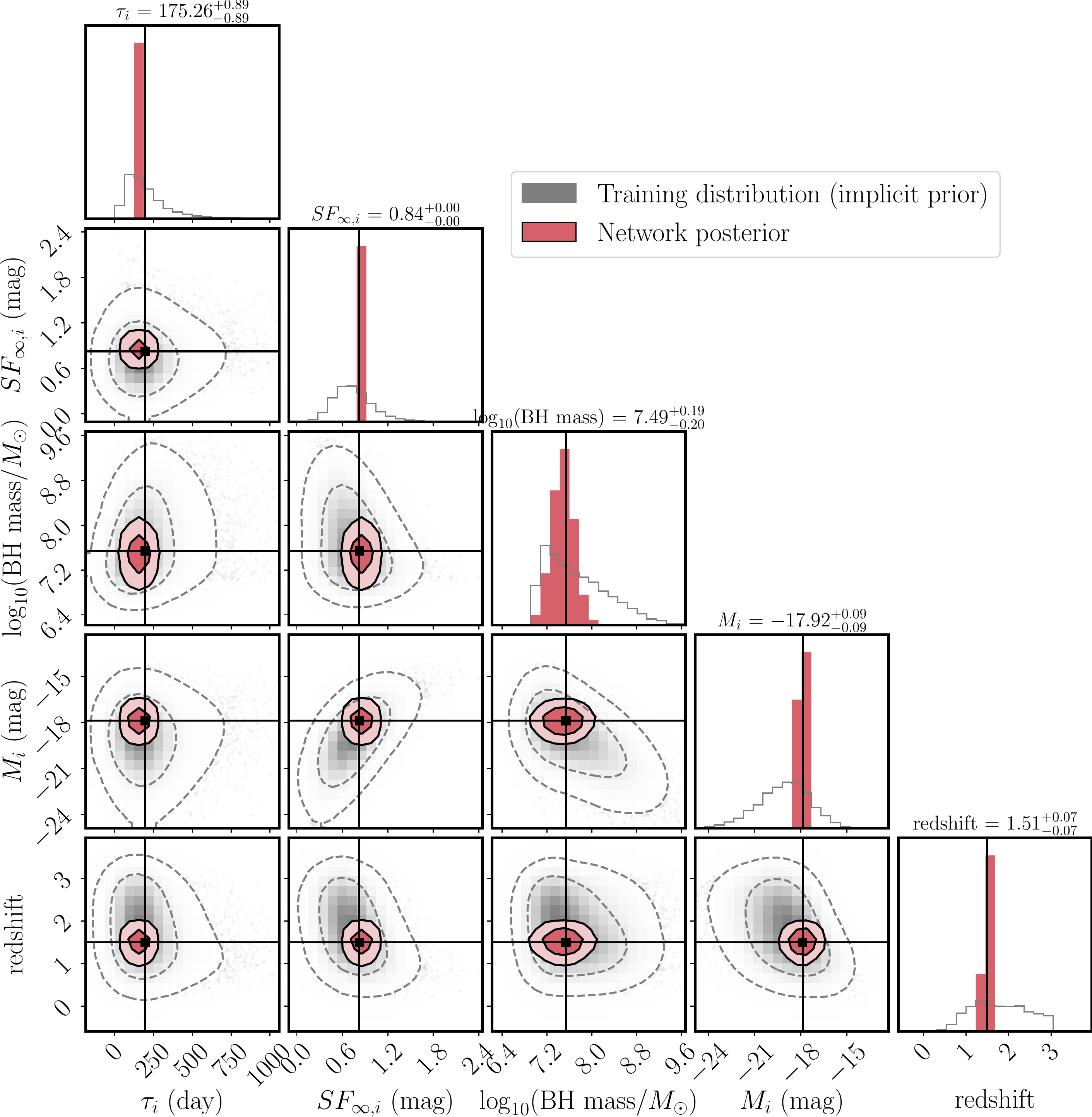}}
\caption{Network posterior for a validation AGN overlaid with the training distribution for a representative subset of the target quantities. The vertical/horizontal black lines show the true parameter values. In this example, our network produces a tight and unbiased estimate for the BH and light curve parameters.}
\label{fig:posterior-corner}
\end{center}
\vskip -0.4in
\end{figure}
Figure \ref{fig:posterior-corner} shows the inferred posterior over a representative subset of the target quantities in $\bm{s}$, for a single AGN. Our BANP can accurately recover the true quantities within 1-$\sigma$ credible interval.

To assess the precision and accuracy of our parameter recovery, we proceed similarly as in Section \ref{sec:results-reconstruction} to evaluate the 1-$\sigma$ uncertainty and MAE of our posterior on the 21 target quantities. The resulting values are listed on Table \ref{metrics}. Our baseline method was a residual MLP that took as input the summary statistics of the light curve---the mean and standard deviation of the flux in six filters---to yield an identically parameterized posterior. We find a twofold improvement in target recovery, in both precision and accuracy. In particular, the timescale-sensitive quantities $\tau$ and BH mass showed the most improvement, of 150\% and 120\%, respectively. The BH mass was precise to 0.4 dex and accurate to 0.3 dex.

\subsubsection{Calibration}
\begin{figure}[ht]
\begin{center}
\centerline{\includegraphics[width=0.9\columnwidth]{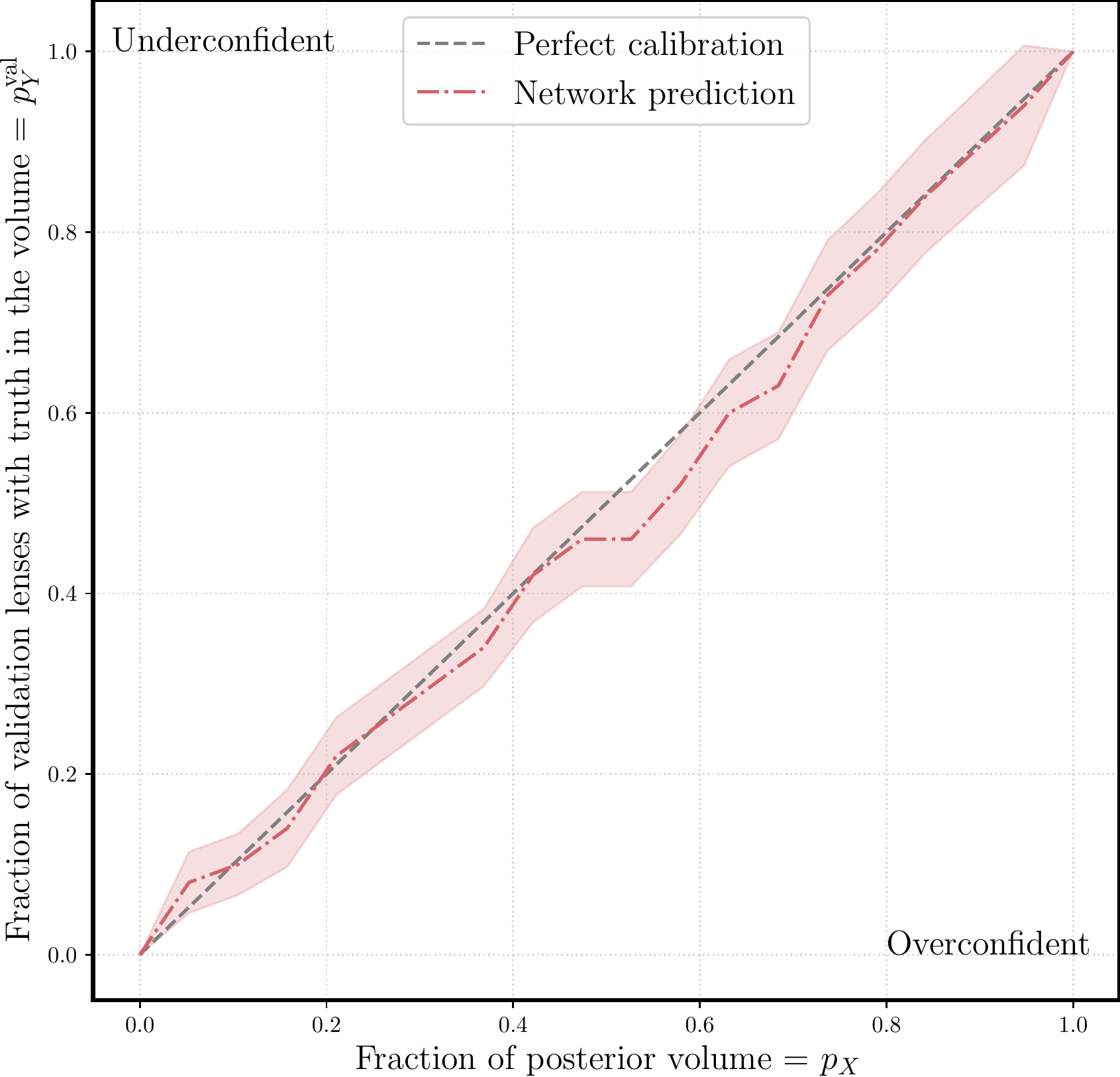}}
\caption{The confidence of our network is statistically consistent with the truth. The calibration metric for our model falls on the $p_Y^{\rm val}=p_X$ line, which indicates perfect calibration. Error bands are estimated using jacknife sampling of the posterior samples.}
\label{fig:calibration}
\end{center}
\vskip -0.4in
\end{figure}

Figure \ref{fig:calibration} plots the calibration metric introduced in Wagner-Carena et al. \citeyearpar{wagner2021hierarchical}.\footnote{The supplementary material describes this metric in detail.} For a given fraction of the BANP posterior probability volume, $p_X$, the metric plots the fraction of posterior samples that contain the truth within the volume, $p_Y$. If the posterior is perfectly calibrated, $p_X$ of the posterior samples would encompass the truth $p_Y = p_X$ of the time, for every value of $p_X$. We apply this metric on the validation set as a whole by averaging the $p_Y$ values across the validation AGN, to get $p_Y^{\rm val}$. Regions of the curve with $p_{\rm Y}^{\rm val}<p_X$ indicate overconfidence whereas the opposite indicates underconfidence. We find good calibration for with our choice of MC dropout rate (0.05).

\section{Discussion and Future Work}
We have adapted latent neural processes for probabilistic parameter estimation and multivariate time series reconstruction. Our method is capable of interpolating AGN light curves and precisely inferring BH parameters. Compared to classical fitting methods, the BANP is flexible and does not assume a fixed parameterization or kernel. This is important for processing real AGN data, because there exists no physical parameterized model that can reliably describe the temporal patterns. Our method is thus useful for extracting an informative latent space from noisy and irregularly sampled time series in general. 

In future work, we plan to upgrade to more physical simulations and asynchronous sampling in $ugrizY$, in preparation for real LSST AGN data. Ultimately, we aspire to hierarchically infer the hyperparameters that govern the AGN population, using the constraints from the individual AGN. Likelihood-free inference methods such as normalizing flows \citep{rezende2015variational} are interesting for a flexible posterior parameterization. We will also explore incorporating physics priors into the network architecture; scalable variants of Gaussian processes (e.g. Salimbeni et al. \citeyear{salimbeni2017doubly}, Wilson et al. \citeyear{wilson2016stochastic}) and latent stochastic differential equations \cite{li2020scalable} are promising alternatives to the ANP for modeling temporal correlations.


\section*{Acknowledgements}


\bibliography{main}
\bibliographystyle{icml2021}

\section*{Supplement}

\subsection*{Calibration metric} \label{sec:calibration}
We summarize the semi-quantitative calibration metric introduced in Wagner-Carena et al. \citeyearpar{wagner2021hierarchical}. This is a multi-dimensional generalization of commonly used confidence-frequency tests (e.g. Niculescu-Mizil et al. \citeyear{niculescu2005predicting}) for evaluating the statistical consistency of model uncertainties. 

Denote the $N$ parameter samples drawn from the Bayesian attentive neural process (BANP) posterior for some AGN $k$ as $\{ \bm{s}_n ^{(k)} \}_{n=1}^{N}$ and the true parameter value as $\bm{s}_{\rm true}^{(k)}$. For a given fraction of the BANP posterior probability volume, $p_X$, the metric queries the fraction of the samples containing the truth within this volume, $p_Y$. More precisely,
\begin{align} \label{eq:p_Y}
    p_{ Y}^{(k)}(p_X) = \mathbbm{1}\left\{ \frac{ \sum_{n=1}^N  \mathbbm{1}\left\{d(\bm{s}_n^{(k)}) < d(\bm{s}_{\rm true}^{(k)})\right\}}{N} < p_X \right\} 
\end{align}
where $\mathbbm{1}\{\cdot \}$ is an indicator function that evaluates to 1 when the argument is true and 0 otherwise, and $d(\bm{s})$ is the distance measure of a particular point $\bm{s}$ from the posterior predictive mean given the posterior width. Following Park et al. \citeyearpar{park2021large}, we use the Mahalanobis distance for $d$. If the posterior is perfectly calibrated, $p_X$ of the samples would encompass the truth $p_Y = p_X$ of the time, for every value of $p_X$. 

This metric can be evaluated on the dataset as a whole by averaging the $p_Y$ values from individual AGN. For $N^{\rm val}$ AGN in the validation set, $p_Y$ and $p_X$ can be expressed as:
\begin{align} \label{eq:p_Y_val}
    p_{ Y}^{\rm val}(p_X) = \frac{1}{N^{\rm val}} \sum_{k=1}^{N^{\rm val}} p_{ Y}^{(k)}(p_X)
\end{align}

Plotting $p_{\rm Y}^{\rm val}$ for a grid of $p_X$ values yields the calibration curve. Regions of the curve with $p_{\rm Y}^{\rm val}<p_X$ indicate overconfidence whereas the opposite indicates underconfidence. 

\end{document}